\documentclass[12pt,preprint]{aastex}
\usepackage{graphics,graphicx,amssymb,amsmath,epsfig,hyperref}
\shorttitle{Building the Halo}
\shortauthors{Maxwell et al.}
\begin{document}
\title{Building the Stellar Halo Through Feedback in Dwarf Galaxies}
\author{Aaron J. Maxwell, James Wadsley, H. M. P. Couchman, \& Sergey
Mashchenko}\affil{Department of Physics \& Astronomy, McMaster University,
Hamilton, ON, L8S 4M1, CAN}\email{ajmax@mcmaster.ca}
\begin{abstract}
We present a new model for the formation of stellar halos in dwarf galaxies.
We demonstrate that the stars and star clusters that form
naturally in the inner regions of dwarfs are expected to
migrate from the gas rich, star forming centre to join the
stellar spheroid.  For dwarf galaxies, this process could be the dominant
source of halo stars.  The effect is caused by stellar feedback-driven
bulk motions of dense gas which, by causing potential fluctuations
in the inner regions of the halo, couple to all collisionless
components.  This effect has been demonstrated to generate cores in
otherwise cuspy cold dark matter profiles and is particularly
effective in dwarf galaxy haloes.  
It can build a stellar spheroid with larger ages and lower
metallicities at greater radii without requiring an outside-in formation model.
Globular cluster-type star clusters can be created
in the galactic ISM and then migrate to the spheroid on 100\thinspace Myr 
timescales.  Once outside the inner regions they are less susceptible to
tidal disruption and are thus long lived; clusters on wider orbits
may be easily unbound from the dwarf to join the halo of a
larger galaxy during a merger.  A simulated dwarf galaxy
($\text{M}_{\rm vir}\simeq10^{9}\text{\thinspace M}_{\odot}$ at $z=5$) is used to
examine this gravitational coupling to dark matter and stars.
\end{abstract}
\section{Introduction}\label{sec:intro}
\indent Dwarf galaxies are the predominant star
forming objects in the early universe and dwarf spheroidals, in
particular, are fossil remnants of this era \citep{dekel1986}.
Normal star formation (post-Population III) occurs in the densest
gas accumulating in the centres of galactic potential wells.
In this case, we might expect dwarf spheroids to be simple, highly concentrated, star piles.  
In contrast,
the stars in observed dwarfs are diffuse and many lack a conspicuous
nucleus.  Further, dwarfs at or above the luminosity of Fornax
have their own globular cluster systems \citep{mateo1998}.  If these galaxies
were the
first objects large enough to have a high-pressure ISM in their centres,
capable of
forming large clusters, we need to explain how such clusters could
end up orbiting at substantial radii with a distribution similar to
that of the overall light \citep{miller2009}.
Radial age and metallicity gradients are also observed \citep{mcconnachie2012},
suggesting an outside-in formation scenario
reminiscent of the ``monolithic collapse'' model \citep[hereafter ELS]{eggen1962}.\\
\indent In this paper, we explore how features present in the old stellar
populations of dwarf galaxies can occur naturally in contemporary cosmological models through star
formation and feedback in these galaxies.
Young dwarf galaxies have a high gas content and form stars
vigorously.  In prior work, \citet{mashchenko2008} were able to show
that stellar feedback in a simulated dwarf galaxy will drive bulk gas
motions that couple gravitationally to all matter near the centre of
the dwarf.  As discussed in \S\ref{fb}, this mechanism has been shown
to act in actively star forming galaxies at a range of masses and is
believed to be generic.  
The process pumps energy into the orbits of all material passing
near the centre, transforming an initial dark matter cusp into a
broad core, consistent with observations.  Here we study the evolution
of the stellar content, which is formed self-consistently in those
simulations.  Orbit pumping also operates on stars, the other key
collisionless component of galaxies, to grow stellar spheroids from
the inside out, as well as place massive star clusters on large radial
orbits.\\
\indent It is widely understood that the $\Lambda$CDM cosmology predicts the
hierarchical assembly of galaxies: dwarf 
proto-galaxies interact and merge into larger galaxies, contrary to the ELS
model.  \citet{searle1978} and \cite{zinn1980} refined this model by invoking 
a late in-fall of old stars that would contribute to both the stellar halo
and its globular cluster population.
Subsequent work has focused on reconciling this picture
for the formation of the Galactic stellar halo with the standard
hierarchical framework \citep[for a recent review see][]{helmi2008}. 
Chemical enrichment models combined with descriptions of a Milky Way
(MW)-type merger history \citep[e.g.][]{robertson2005,bullock2005,delucia2008}
and cosmological simulations of MW-type galaxies
\citep[e.g.][]{zolotov2010} 
can be made to match the abundance
patterns of the stellar halo \citep[e.g.][]{carollo2007,carollo2010,dejong2010}.
A general conclusion is that dwarf progenitors play a
major role in building the MW halo, owing to their high rates of star
formation at early times and their ability
to retain supernova-enriched gas.\\
\indent However, MW-scale simulations poorly
resolve dwarf galaxies which thus readily disintegrate and contribute 
their entire stellar contents to the halo. This conclusion is a direct
consequence of low numerical resolution and is at odds with how star
formation would be expected to occur in dwarfs.  
A closer understanding of star formation in dwarf galaxies is needed to establish
how those stars are produced and how readily they can contribute to the observed Galactic
stellar populations and their radial variations.\\
\indent In \S\ref{fb} we discuss the stellar redistribution mechanism and how it operates.
In \S\ref{stars} we explore the impact this has on the formation of
the stellar spheroid in dwarfs including stellar systems such as globular
clusters.  We also discuss implications for dwarfs contributing their
stars to the spheroids of larger galaxies.
\section{Dynamical Impact of Stellar Feedback}\label{fb}
Observations of the kinematics of the stellar and gaseous components
of dwarf galaxies point to these systems having a cored density profile \citep[e.g.][see
\citet{deblok2010} for a recent review]
{burkert1995,cote2000,gilmore2007,oh2011a} in contrast to
collisionless simulations of Cold Dark Matter (CDM) haloes which
predict a central density cusp
\citep[e.g.][]{dubinski1991,navarro1995,bullock2001,
  klypin2001,stadel2009}.  \citet{mashchenko2008} presented a
solution to this long-standing challenge for the CDM cosmogony by
correctly accounting for the 
the impact of stellar feedback.  By feeding the energy generated by supernovae into the
surrounding star-forming gas, they were able to generate fluctuations in the
gravitational potential that pumped the dark matter orbits and removed
the cusp. The effectiveness
of dark matter orbit pumping due to stellar feedback has been confirmed in simulations by
other workers, showing that it operates in dwarfs to the present day \citep{governato2010}
and also affects larger galaxies up to Milky Way scales with sufficiently strong
feedback \citep{maccio2012}.  In addition to operating in the SPH
code used by \citet{mashchenko2008}, the mechanism has also
been demonstrated using a grid code\footnote{R. Teyssier, priv. comm.}.\\
\indent Two critical features allowed \citet{mashchenko2008} to
demonstrate the effect of stellar feedback on dark matter orbits in
the dwarf galaxy: high resolution 
($300\text{\thinspace M}_{\odot}$ per gas particle) and low temperature metal
cooling (10--$8000\text{\thinspace K})$.  The combination of these features
allowed the formation of a cold, dense gas phase and permitted the use
of a far more realistic minimum density for star formation,
$\sim100\text{ atoms cm}^{-3}$, comparable to molecular cloud
densities.  This was in sharp contrast to prior work where star
formation occurred fairly uniformly throughout the ISM of simulated
galaxies.  As a result this was the first cosmological simulation to
form numerically resolved star clusters up to $\sim10^{5}\text{\thinspace
  M}_{\odot}$.\\
\indent A direct result of clustered star formation is highly
localized and episodic feedback that violently rearranges the gas in
the inner regions of the dwarf galaxy. Since the gas dominates the
mass in the star forming region, this results in a gravitational
potential that varies on a timescale commensurate with orbital times.
This
causes irreversible changes in the orbital energies of all
matter passing near the star-forming centre of the dwarf. Whereas sharp changes
in the potential impulsively modify all particle orbits \citep{pontzen2012},
\cite{mashchenko2006} showed that oscillating potentials with speeds
closer to the typical particle velocity couple strongly and flatten the
core more rapidly (their figure 2).  For material that initially has a
low velocity dispersion, such as dark matter within the cusp, this
preferentially increases the orbital radius and redistributes the
material into a smooth core as shown in \cite{mashchenko2008} and
other works. For the gaseous component, shocks dissipate this
added velocity whereas dark matter and stars undergo a random walk in orbital energy.\\
\begin{figure}
\centering
\includegraphics[width=12cm,height=10cm]{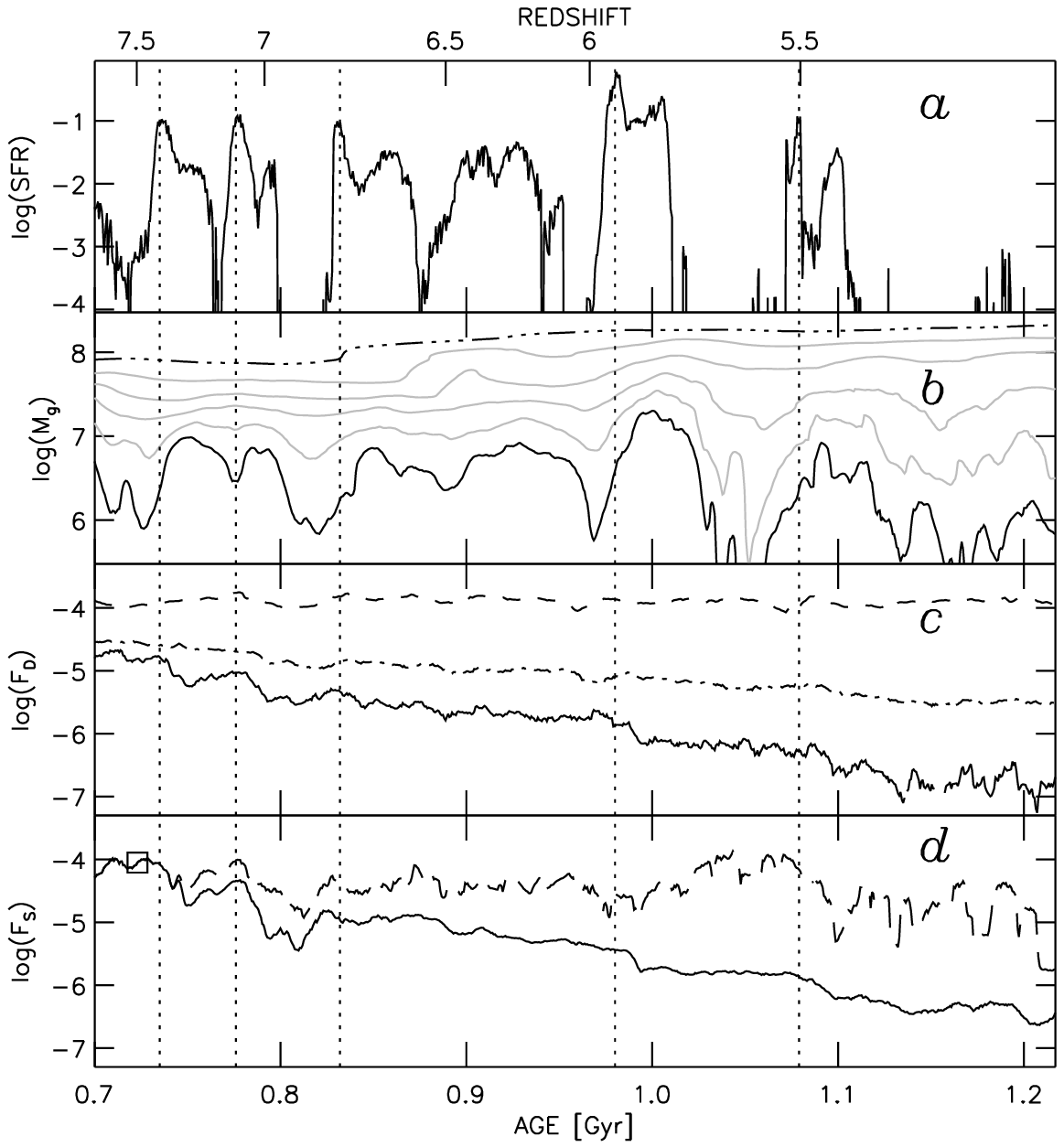} 
\caption[]{Star formation rate, variation of enclosed gas mass and
  variation of phase-space density for the collisionless components.\\
\emph{a}) Star formation rate (M$_{\odot}\text{\thinspace yr}^{-1}$)
within a radius of 100\thinspace pc of the centre.  Vertical dotted lines
highlight 5 strong star formation peaks.\\
\emph{b}) Enclosed gas mass (M$_{\odot}$) for various radii.  The solid line 
is for a 100\thinspace pc radius, and the grey lines increase the radius by a
factor of two each step.  The triple-dot-dash line is the enclosed mass within 3.2\thinspace kpc.\\
\emph{c}) The dark matter phase-space density (M$_{\odot}\text{\thinspace
pc}^{-3}\text{\thinspace km}^{-3}\text{\thinspace s}^{3}$) measured within 100\thinspace pc.  The
solid line indicates low velocity dark matter (see text) while the 
dash-dot is for all dark matter.  The dashed line shows the behaviour
in a simulation
without stellar feedback.\\
\emph{d}) The stellar phase-space density
(M$_{\odot}\text{\thinspace pc}^{-3}\text{\thinspace km}^{-3}\text{\thinspace s}^{3}$)
measured within 100\thinspace pc.  The solid line is for stars formed before $z=7.5$ (time stamp denoted by the square), while
the long-dash line is for all stars.}
\label{fig:1}
\end{figure}
\indent We use the simulated dwarf of
\cite{mashchenko2008} to illustrate the process.  We selected a period
between redshifts 8--5 without major mergers so that the evolution is
dominated by centralized star formation fueled by a consistent gas supply.
Figure \ref{fig:1} shows
several cycles of star formation, feedback upon the gas content and
the response of the collisionless components.  In this simulation,
the majority of stars form within 100\thinspace pc, which we use as a radial
size in which to measure the feedback effects. The centre is defined as the
position of the 100 most bound particles. This choice biases towards
gas-rich star-forming regions but gives very similar results
to using a mass-weighted centre. The central 100\thinspace pc region is well
resolved in space and mass.\\
\indent In Figure \ref{fig:1}a, the total star formation rate in units of
M$_{\odot}\text{\thinspace yr}^{-1}$ is shown for all stars formed within 100\thinspace pc.
The star formation is highly episodic in this redshift range.  Given
that star formation is confined to a small central region, stellar
feedback is very efficient at cutting off the supply of cold, dense
gas that fuels the process.  In the feedback model employed in this
simulation, the effective component is supernova energy injection
acting over a period of 10--30\thinspace Myr after initial star formation.
Thus, once initiated within a dense knot of gas, star formation rises
to a peak and shuts down in around 10\thinspace Myr.
The five highest peaks in the central star formation rate are marked with
vertical dotted lines to guide the eye over the four panels of the figure.\\
\indent The solid line in Figure \ref{fig:1}b shows the enclosed gas mass in
units of M$_{\odot}$ as a function of time within 100\thinspace pc.  The enclosed
gas mass shows the same cyclic behaviour as the star formation rate
with a lag of 10--20\thinspace Myr.  Gas falls into the inner regions, forming
dense clouds and allowing star formation to begin.  Stellar feedback starts
to pressurize the gas leading to both compression and the
driving of material out of the inner regions.  The gas velocity dispersion
varies from 10--$40\text{\thinspace km\thinspace s}^{-1}$ within 100\thinspace pc, indicating crossing
times of roughly 5--20\thinspace Myr.  Thus the gas mass peaks slightly after the
peak in star formation and then subsides.\\
\indent The total gas mass (triple-dot-dash line in the same panel)
within 3.2\thinspace kpc (the virial radius at $z=8$) increases steadily
due to fresh in-falling material, reaching $\sim2\times10^{8}\text{\thinspace M}_{\odot}$
at $z=5$.  Feedback associated with vigorous star formation can readily create
hot gas ($\text{T}>10^{6}\text{\thinspace K}$) and outflows exceeding the
$100\text{\thinspace km\thinspace s}^{-1}$ escape velocity.
Such unbound gas can travel tens of kiloparsecs from the dwarf.  However, the
total mass in unbound (mostly hot) gas generated is comparable to the
$3\times10^{7}\text{\thinspace M}_{\odot}$ in stars created over the 500\thinspace Myr
period shown in Figure \ref{fig:1}.  In-fall onto the galaxy continues steadily
along cold filaments next to the outflow channels and is not disrupted by the
outflow, as the figure indicates.  The baryon fraction inside the
virial radius is always moderately in excess of the universal baryon
fraction.  The gas mass within the star forming inner region
fluctuates dramatically in response to feedback but much of this gas
is simply cycling within the inner few hundred parsecs.  Within 800\thinspace pc the gas
content grows fairly smoothly as shown by the second-to-top grey curve.\\
\indent The numerical values for star formation rates and mass
outflows are a result of the specific sub-grid models and resolution
used for this simulation (though the resolution is much higher than is
typical).  However, the qualitative picture is expected to be robust and
is consistent with our understanding of feedback and its role in
creating a bursty star formation history in smaller galaxies \citep{stinson2007}.
The gas within the entire halo is characterized by churning motions with colder
gas moving in and hotter gas moving out.  This is in contrast to the
simple gas blow-out picture of the evolution of dwarf galaxies
\citep[e.g][]{navarro1996b} where the entire star-forming gas content is at
least temporarily evacuated.  The advantage here is the continual
availability of gas for ongoing star formation with bursts on the
dynamical timescale of the dense inner regions (50--100\thinspace Myr) that repeatedly
perturb the collisionless components.  This type of churning also occurs in
more massive galaxies \citep{brook2012}.\\
\indent Figure \ref{fig:1}c shows the behaviour of the central
dark matter phase-space density, approximated as $\rho/\sigma^{3}$, where
$\rho$ is the mean dark matter density and $\sigma$ is the velocity
dispersion within the central 100\thinspace pc. Whereas the fine-grained, dark matter,
6-dimensional phase-space density is strictly conserved, the 
coarse-grained dark matter phase-space density, which we are
approximating with $\rho/\sigma^{3}$, is insensitive to
adiabatic compression due to baryonic dissipation but should decrease 
monotonically due to irreversible (non-adiabatic) effects.\\ 
\indent The dot-dash line is for all dark matter particles located within 100\thinspace pc
of the centre at each simulation output.  The solid line shows the behaviour of
a group of dark matter particles with velocities less than
$20\text{\thinspace km\thinspace s}^{-1}$,
selected at $z=8$, and followed throughout the simulation.
Both dark matter groups show a steady decrease in the coarse-grained
phase-space density 10--20\thinspace Myr after a star formation peak.
Note, however, that the low velocity dark matter exhibit much steeper gradients.
This decrease is associated with both an increase
in their velocity dispersion and a decrease in their density.  This
persistent decrease of $\rho/\sigma^{3}$, unaffected by the gas returning to
the core, demonstrates that the heating of collisonless matter through the
gas-driven gravitational potential oscillations is irreversible.
The long term effect is a random
walk in orbital energy that redistributes dark matter particles into a
cored profile of order 400\thinspace pc in size by $z=5$ as shown by
\citet{mashchenko2008}.
In
the same dwarf simulated without any star formation, the phase-space
density remains relatively constant as shown by the dashed line.
\section{Stars: The Other Collisionless Component}\label{stars}
Stars also behave as a collisionless fluid, and so couple to the
potential fluctuations created by stellar-feedback-driven gas motions.
Indeed, the majority of stars are formed within the dwarf core and
spread outwards by the end of the simulation.  The central stellar
density is regularly increased by new stars which are then dispersed
to larger orbits.  The long-dash line in Figure
\ref{fig:1}d shows the phase-space density for all stars within 100\thinspace
pc.  As the stellar density within this radius is 
replenished by star formation, there is no significant trend in the
phase-space density.  To examine the evolution of stars after their
formation, we selected stars within the inner 100\thinspace pc that were
formed before 0.72\thinspace Gyr and tracked their phase-space density over time
as indicated by the solid line.  The decrease in phase-space density
is dramatic, comparable to that 
for the low-velocity-dispersion dark matter.  Since the stars form from gas
with velocity dispersion $\sim 20\text{\thinspace km\thinspace s}^{-1}$,  these trends
reflect the greater efficiency of this heating mechanism
for low velocity material.  Note also that significant decreases in
the phase-space density occur 10--20\thinspace Myr after the central bursts of
star formation, as discussed above.  In our simulated dwarf, stellar
orbits are found to expand to beyond 1\thinspace kpc through the effects of this
mechanism.
\subsection{The Diffuse Spheroid}
\begin{figure}
\centering
\includegraphics[width=8.5cm]{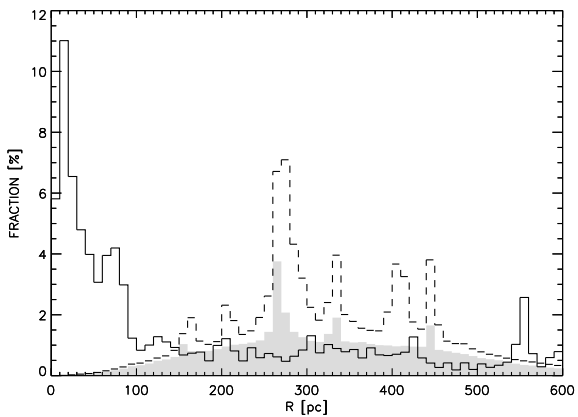} 
\caption[]{The normalized distribution of stellar formation radius (solid) and the
final stellar radius (dashed).  The shaded region corresponds to the distribution
of final radii for the stars that formed within 100\thinspace pc.}
\label{fig:2}
\end{figure}

In this new picture, stars form preferentially in spatially concentrated star
bursts near the gas-rich centres of small galaxies and
then migrate to the outer parts of the galaxy.  Orbital 
changes occur repeatedly for objects traversing the star forming core.  
This effect
will not be limited to small galaxies but may become less pronounced
for larger halos.  The degree to which stars have migrated is a
function of their time of formation and the period of time for which sufficiently vigorous
potential fluctuations were available to pump their orbits.  This
provides an alternative to simply scaling the ELS view down to smaller halo
masses.\\
\indent Figure \ref{fig:2} shows that even though half the stars formed inside
100\thinspace pc (solid line), by the end of the simulation (500\thinspace Myr later) they
fill the entire dwarf halo with no distinction between those that
formed inside and outside 100\thinspace pc.   New stars take time to move outward
and when star formation stops, so does the orbital expansion.  The
result at $z=5$ is a moderate trend to larger stellar ages with
radius. This may explain the age and metallicity gradients believed to be
present in the Local Group Dwarfs \citep{mateo1998}.  
\subsection{Bound Star Clusters}
\begin{figure}
\centering
\includegraphics[width=8.5cm]{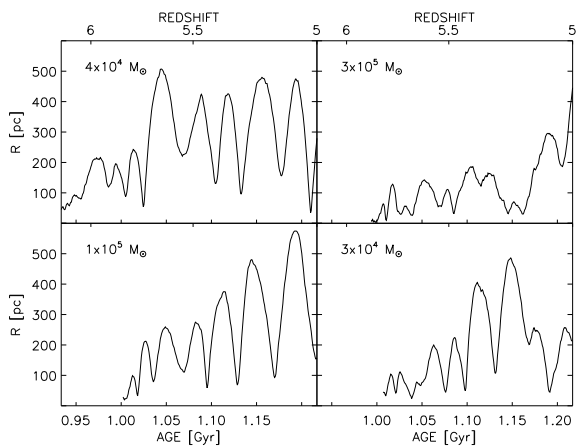} 
\caption[]{Evolution of orbital radii for four long-lived star clusters in the simulation.}
\label{fig:3}
\end{figure}

As noted above, the star formation that occurs in the simulation is
clustered in character due to the unusually high spatial resolution
($100\text{\thinspace M}_{\odot}$ per star particle) and modeling of low temperature cooling \citep{mashchenko2008},
consistent with the majority of star formation in nature.  The majority of these clusters
are disrupted as the simulation progresses, and the stars are deposited
across the stellar spheroid of the dwarf.  This is partly a resolution
effect as the gravitational resolution of the simulation (10\thinspace pc) will
not result in smaller, more tightly bound
clusters.   There are, however, a few of these 
clusters that survived for at least 200\thinspace Myr.
The four most massive and well
resolved clusters (100--1000 stellar particles) were identified within the dwarf
spheroid near the end of the simulation and their orbits traced backwards to the point at which 10\% of the stars within
each cluster had formed. The radial component of the orbits for the
four clusters are shown in Figure \ref{fig:3}.
These four massive clusters form well within 100\thinspace pc but are then driven
out to large radii as each pericentric passage brings them close to the
actively star-forming galactic centre.  The process is a random walk with an average tendency
to increase the apocentric distance.  These clusters also show changes in the
direction and magnitude of their orbital angular momenta.  The process should
be less effective for higher orbital velocities and is thus expected to
saturate when the orbits are well outside the star forming region.\\
\indent Our approach to the building of dwarf spheroids may shed light on the
formation of Globular Clusters. Since the same mechanism migrates both stars
and stellar clusters to the diffuse spheroid, it provides a natural explanation for the
radial distribution seen in dwarf galaxies outside the Local Group
\citep[e.g.][]{miller2009}.  Furthermore, the time that the clusters are resident
in the inner star forming region (which has grown to $\sim 300\text{\thinspace pc}$ by $z=5$) is typically
at least 10$^{8}$\thinspace yr.  Visual inspection of the simulation indicates that dense
gas knots move with the clusters during this period, thus providing a
simple explanation for
the recently observed multiple generations of stars within globular
clusters \citep[e.g][]{dercole2010}.
\section{Conclusions}
We have presented a new framework for understanding the formation of the
stellar spheroid in dwarf galaxies: All stars form in the nuclear
regions and are then redistributed to eventually occupy the entire
halo.  The redistribution mechanism relies on strong fluctuations in
the baryon-dominated central gravitational potential that are associated with stellar
feedback as first demonstrated by \cite{mashchenko2008}.  These fluctuations
irreversibly affect the orbits and hence distributions of the collisionless
components: dark matter, stars and star
clusters.  The key implications are:\\[10pt]
\noindent $\bullet\ $ This process directly affects dwarf
  galaxies.  In these galaxies a mild gradient with radius of increasing age and
  decreasing metallicity would be created as older stars achieve the
  largest orbits.  Orbital redistribution stops when vigorous star
  formation ceases.\\[10pt]
\noindent $\bullet\ $ The central density of stars stays fairly constant as new stars
  form to replace those migrating outwards.\\[10pt]
\noindent $\bullet\ $ Globular cluster-like star clusters form in the ISM (and thus
  have no associated dark matter) and migrate outward over several
  orbital periods.\\[10pt]
\noindent $\bullet\ $ The star clusters may form multiple generations of
  stars from enriched gas readily available in the nuclear regions.  They will
  lose access to new gas as their orbits become larger.\\[10pt]
\noindent $\bullet\ $ Continuous creation and outward migration of
stars and globular clusters avoids the formation of a super-nucleus at
the centre of most dwarf galaxies.\\[10pt]
\noindent $\bullet\ $ Larger clusters become protected against tidal
destruction as their orbits grow and the dwarf's dark-matter core
becomes flattened.\\[10pt]
\noindent $\bullet\ $ Mergers and tidal stripping will deposit these loosely bound
  stars and clusters into the halo of later generations of larger
  galaxies.\\[10pt]
\noindent $\bullet\ $ Large star clusters formed in dwarf galaxies at high redshift,
  rather than in dark matter mini-halos, could be the primary source of
  Globular Clusters in all galaxies.
\acknowledgements This work has made use of the SHARCNET Consortium,
part of the Compute/Calcul Cananda Network.  AJM, HMPC and JWW are
recipients of NSERC funding. HMPC achowledges support from CIfAR.


\begin{thebibliography}{34}
\expandafter\ifx\csname natexlab\endcsname\relax\def\natexlab#1{#1}\fi
\bibitem[{{Brook} {et~al.}(2012){Brook}, {Stinson}, {Gibson}, {Ro{\v s}kar},{Wadsley}, \& {Quinn}}]{brook2012}
{Brook}, C.~B., {Stinson}, G., {Gibson}, B.~K., et al. 2012, \mnras, 419, 771
\bibitem[{{Bullock} \& {Johnston}(2005)}]{bullock2005}
{Bullock}, J.~S. \& {Johnston}, K.~V. 2005, \apj, 635, 931
\bibitem[{{Bullock} {et~al.}(2001){Bullock}, {Kolatt}, {Sigad}, {Somerville}, {Kravtsov}, {Klypin}, {Primack}, \& {Dekel}}]{bullock2001}
{Bullock}, J.~S., {Kolatt}, T.~S., {Sigad}, Y., et al. 2001, \mnras, 321, 559
\bibitem[{{Burkert}(1995)}]{burkert1995}
{Burkert}, A. 1995, \apjl, 447, L25
\bibitem[{{Carollo} {et~al.}(2010){Carollo}, {Beers}, {Chiba}, {Norris}, {Freeman}, {Lee}, {Ivezi{\'c}}, {Rockosi}, \& {Yanny}}]{carollo2010}
{Carollo}, D., {Beers}, T.~C., {Chiba}, M., et al. 2010, \apj, 712, 692
\bibitem[{{Carollo} {et~al.}(2007){Carollo}, {Beers}, {Lee}, {Chiba}, {Norris}, {Wilhelm}, {Sivarani}, {Marsteller}, {Munn}, {Bailer-Jones}, {Fiorentin}, \& {York}}]{carollo2007}
{Carollo}, D., {Beers}, T.~C., {Lee}, Y.~S., et al. 2007, \nat, 450, 1020
\bibitem[{{C{\^o}t{\'e}} {et~al.}(2000){C{\^o}t{\'e}}, {Carignan}, \& {Freeman}}]{cote2000}
{C{\^o}t{\'e}}, S., {Carignan}, C., \& {Freeman}, K.~C. 2000, \aj, 120, 3027
\bibitem[{{de Blok}(2010)}]{deblok2010}
{de Blok}, W.~J.~G. 2010, Advances in Astronomy, 2010
\bibitem[{{de Jong} {et~al.}(2010){de Jong}, {Yanny}, {Rix}, {Dolphin}, {Martin}, \& {Beers}}]{dejong2010}
{de Jong}, J.~T.~A., {Yanny}, B., {Rix}, H.-W., et al. 2010, \apj, 714, 663
\bibitem[{{De Lucia} \& {Helmi}(2008)}]{delucia2008}
{De Lucia}, G. \& {Helmi}, A. 2008, \mnras, 391, 14
\bibitem[{{Dekel} \& {Silk}(1986)}]{dekel1986}
{Dekel}, A. \& {Silk}, J. 1986, \apj, 303, 39
\bibitem[{{D'Ercole} {et~al.}(2010){D'Ercole}, {D'Antona}, {Ventura}, {Vesperini}, \& {McMillan}}]{dercole2010}
{D'Ercole}, A., {D'Antona}, F., {Ventura}, P., {Vesperini}, E., \& {McMillan}, S.~L.~W. 2010, \mnras, 407, 854
\bibitem[{{Dubinski} \& {Carlberg}(1991)}]{dubinski1991}
{Dubinski}, J. \& {Carlberg}, R.~G. 1991, \apj, 378, 496
\bibitem[{{Eggen} {et~al.}(1962){Eggen}, {Lynden-Bell}, \& {Sandage}}]{eggen1962}
{Eggen}, O.~J., {Lynden-Bell}, D., \& {Sandage}, A.~R. 1962, \apj, 136, 748
\bibitem[{{Gilmore} {et~al.}(2007){Gilmore}, {Wilkinson}, {Wyse}, {Kleyna}, {Koch}, {Evans}, \& {Grebel}}]{gilmore2007}
{Gilmore}, G., {Wilkinson}, M.~I., {Wyse}, R.~F.~G., et al. 2007, \apj, 663, 948
\bibitem[{{Governato} {et~al.}(2010){Governato}, {Brook}, {Mayer}, {Brooks}, {Rhee}, {Wadsley}, {Jonsson}, {Willman}, {Stinson}, {Quinn}, \& {Madau}}]{governato2010}
{Governato}, F., {Brook}, C., {Mayer}, L., et al. 2010, \nat, 463, 203
\bibitem[{{Helmi}(2008)}]{helmi2008}
{Helmi}, A. 2008, \aapr, 15, 145
\bibitem[{{Klypin} {et~al.}(2001){Klypin}, {Kravtsov}, {Bullock}, \& {Primack}}]{klypin2001}
{Klypin}, A., {Kravtsov}, A.~V., {Bullock}, J.~S., \& {Primack}, J.~R. 2001, \apj, 554, 903
\bibitem[{{Macci{\`o}} {et~al.}(2012){Macci{\`o}}, {Stinson}, {Brook}, {Wadsley}, {Couchman}, {Shen}, {Gibson}, \& {Quinn}}]{maccio2012}
{Macci{\`o}}, A.~V., {Stinson}, G., {Brook}, C.~B., et al. 2012, \apjl, 744, L9
\bibitem[{{Mashchenko} {et~al.}(2006){Mashchenko}, {Couchman}, \& {Wadsley}}]{mashchenko2006}
{Mashchenko}, S., {Couchman}, H.~M.~P., \& {Wadsley}, J. 2006, \nat, 442, 539
\bibitem[{{Mashchenko} {et~al.}(2008){Mashchenko}, {Wadsley}, \& {Couchman}}]{mashchenko2008}
{Mashchenko}, S., {Wadsley}, J., \& {Couchman}, H.~M.~P. 2008, Science, 319, 174
\bibitem[{{Mateo}(1998)}]{mateo1998}
{Mateo}, M.~L. 1998, \araa, 36, 435
\bibitem[{{McConnachie}(2012)}]{mcconnachie2012}
{McConnachie}, A.~W. 2012, \aj, 144, 4
\bibitem[{{Miller}(2009)}]{miller2009}
{Miller}, B.~W. {Globular Clusters in Dwarf Galaxies}, ed. {Richtler, T.~\& Larsen, S.}, 141
\bibitem[{{Navarro} {et~al.}(1996){Navarro}, {Eke}, \& {Frenk}}]{navarro1996b}
{Navarro}, J.~F., {Eke}, V.~R., \& {Frenk}, C.~S. 1996, \mnras, 283, L72
\bibitem[{{Navarro} {et~al.}(1995){Navarro}, {Frenk}, \& {White}}]{navarro1995}
{Navarro}, J.~F., {Frenk}, C.~S., \& {White}, S.~D.~M. 1995, \mnras, 275, 720
\bibitem[{{Oh} {et~al.}(2011){Oh}, {de Blok}, {Brinks}, {Walter}, \& {Kennicutt}}]{oh2011a}
{Oh}, S.-H., {de Blok}, W.~J.~G., {Brinks}, E., {Walter}, F., \& {Kennicutt}, Jr., R.~C. 2011, \aj, 141, 193
\bibitem[{{Pontzen} \& {Governato}(2012)}]{pontzen2012}
{Pontzen}, A. \& {Governato}, F. 2012, \mnras, 421, 3464
\bibitem[{{Robertson} {et~al.}(2005){Robertson}, {Bullock}, {Font}, {Johnston}, \& {Hernquist}}]{robertson2005}
{Robertson}, B., {Bullock}, J.~S., {Font}, A.~S., {Johnston}, K.~V., \& {Hernquist}, L. 2005, \apj, 632, 872
\bibitem[{{Searle} \& {Zinn}(1978)}]{searle1978}
{Searle}, L. \& {Zinn}, R. 1978, \apj, 225, 357
\bibitem[{{Stadel} {et~al.}(2009){Stadel}, {Potter}, {Moore}, {Diemand}, {Madau}, {Zemp}, {Kuhlen}, \& {Quilis}}]{stadel2009}
{Stadel}, J., {Potter}, D., {Moore}, B., et al. 2009, \mnras, 398, L21
\bibitem[{{Stinson} {et~al.}(2007){Stinson}, {Dalcanton}, {Quinn}, {Kaufmann}, \& {Wadsley}}]{stinson2007}
{Stinson}, G.~S., {Dalcanton}, J.~J., {Quinn}, T., {Kaufmann}, T., \& {Wadsley}, J. 2007, \apj, 667, 170
\bibitem[{{Zinn}(1980)}]{zinn1980}
{Zinn}, R. 1980, \apj, 241, 602
\bibitem[{{Zolotov} {et~al.}(2010){Zolotov}, {Willman}, {Brooks}, {Governato}, {Hogg}, {Shen}, \& {Wadsley}}]{zolotov2010}
{Zolotov}, A., {Willman}, B., {Brooks}, A.~M., et al. 2010, \apj, 721, 738
\end{thebibliography}
\end{document}